\documentclass[useAMS,usenatbib]{mn2e}
\usepackage{graphicx,graphics,amsmath}
\usepackage{latexsym}
\usepackage{float}
\usepackage[caption = false]{subfig}

\onecolumn
\usepackage{xcolor}

\begin{document}
\title[GW from PT of NS to QS]{Gravitational wave signature from phase transition of a combusting neutron star to quark star}

\author[Ritam Mallick \& Shailendra Singh \& R Prasad]
{Ritam Mallick$^{1}$\thanks{mallick@iiserb.ac.in}, Shailendra Singh$^{1}$\thanks{shailendra17@iiserb.ac.in},  \& R Prasad$^{1}$\thanks{rprasad@iiserb.ac.in}\\
	$^{1}$ Indian Institute of Science Education and Research Bhopal, Bhopal, India}
 
\maketitle

\begin{abstract}
Fluctuation at the neutron star center gives rise to a small deconfined quark core very close to the star center. The density discontinuity at the quark-hadron boundary initiates a shock wave, which propagates outwards of the star. The shock has enough energy to combust nuclear matter to 2-flavor quark matter in the star. The 2-flavor quark matter is not stable and settles to a stable 3-flavor matter in the weakly interacting timescale. In this paper, we study the conversion of 2-flavor matter to 3-flavor matter. We set up a differential equation to convert the excess of down quarks to strange quarks involving weak reaction and diffusion of quarks. Calculating the reaction rate, we solve the differential equation to find the velocity of the conversion front. As the conversion front moves out, the density profile changes, bringing about a change in the star's quadrupole moment and thereby emitting gravitational waves. We find that the GW signal depends strongly on the star temperature and mass. The GW amplitude of a colder star is well within present detector capability, but the frequency is slightly on the higher side. Relatively hotter stars are on the boundary of present detectors and easily detectable with future detectors, and their frequency is also within the present detectability range. In comparison, PT from galactic pulsars is easily detectable with present detectors.
\end{abstract}
	
\section{Introduction}

The theory of strongly interacting particles, known as Quantum chromodynamics, predicts that at high temperature and/or density, the fundamental degree of freedom is not hadrons but quarks and gluons. Therefore, at high enough density/temperature, there is a deconfinement transition from hadrons to quarks and gluons. Earth-based experiments like LHC, RHIC probes the matter at high temperatures and non-vanishing small chemical potential. On the other hand, estimating the matter properties at finite chemical potential can only be probed through astrophysical observation from neutron stars (NSs).
Simultaneous detection of gravitational waves (GW) and electromagnetic signals from binary neutron star mergers (BNSM) has given hope that we can probe NS's interior with better precision and improve the understanding of the matter properties occurring at high densities.

The astrophysical signal from the BNSM GW170817 \citep{abbott} was detected from the inspiral phase, from which the tidal deformability of the inspiral stars could be
extracted and thereby the estimates of their masses and spins \citep{faber,baiotti}. The merger and the postmerger phase of the binary collision are still beyond the scope of aLIGO where one expects the density and temperature to increase and thereby the chances of phase transition (PT) from nuclear matter (NM) to quark matter (QM) \citep{hinderer,read,pozzo,agathos,chatzi}. 
Recently Most et al. \citep{most} did a binary merger simulation and found that there would be a phase difference in the evolution of the GW signal during the postmerger phase if the hypermassive NS has quarks in them. Also, it was shown that the empirical relationship between postmerger frequency and tidal deformability deviates significantly for hypermassive stars, which have undergone PT during merging \citep{bauswein}. Therefore, such signals' holy grail remains in the merging and the postmerger phase, which is still beyond the present detectors' capability. 

In most of the studies, the actual process of PT is not captured and is assumed to happen instantaneously. However, if the PT is shock-induced, then the actual process of PT can leave imprints which can be detected with GW detectors. The shock-induced PT has been studied previously by several authors \citep{abhijit,drago,mishustin,prasad,prasad1,irfan}. Primarily they discuss that a sudden density fluctuation in the star induces a shock wave at the star core. It can be the star settling into a stable configuration during mass accretion or after a supernova. This shock is strong enough to combust NM to QM, and as the shock wave propagates out, it mimics PT in the star. The PT is assumed to be a two-step process; the first step converts NM to deconfined 2-flavor (2-f) QM (having only up (u) and down (d) quarks) in the strongly interacting timescale, and the second step converts unstable 2-f QM to stable 3-flavor (3-f) QM (comprising up, down and strange (s) quarks) in the weak timescale \citep{olinto,abhijit,jaikumar,alford15,drago15}. Olinto \citep{olinto} and later by Bhattacharyya et al. \citep{abhijit} studied the weak decay conversion process for low temperature, assuming the reaction rate to be independent of temperature and calculated the velocity of the conversion front and total time taken for the conversion process for different star (ranging from proto NS to cold NS). Later calculations \citep{jaikumar,alford15,drago15} involving time-dependent reaction rate were also carried out where it was found that the conversion process is significantly affected by the temperature of the star. 
	The community is also divided on whether the conversion process is fast detonation or a slow deflagration. Bhattacharyya et al. \citep{abhijit} assumed it to be fast combustion induced by a shock wave. However, other possibilities also exist like deflagration \citep{drago,herzog}. Drago et al. \citep{drago} estimated the energy released from the conversion process and calculated the temperature rise of the star, and concluded that the conversion process is always a deflagration. They concluded that hydrodynamic instabilities could develop, which may increase the conversion velocity. A 3-D numerical simulation \citep{herzog} showed that a turbulent deflagration conversion process could make the front velocity quite fast. 
	
	This article assumes a shock-induced phase transition and mainly studies the weak conversion of the excess down to strange quarks. Our basic calculation is done for cold NS (temperature of the order of $10^{-2}$ MeV); however, we have also estimated our result for higher temperature for comparison. The paper is arranged as follows. Section II discusses the formalism of how we model the 2-f to 3-f combustion and what is the velocity of the combustion front. Section III discussed the calculation and results of the GW signals and the general template of such PT. Finally, in Section IV, we discuss our results and conclude from them.	
	
\section{2-f to 3-f conversion}

\begin{figure}
	%\vskip 0.2in
	\centering
	\includegraphics[width = 3.2in,height=3.0in]{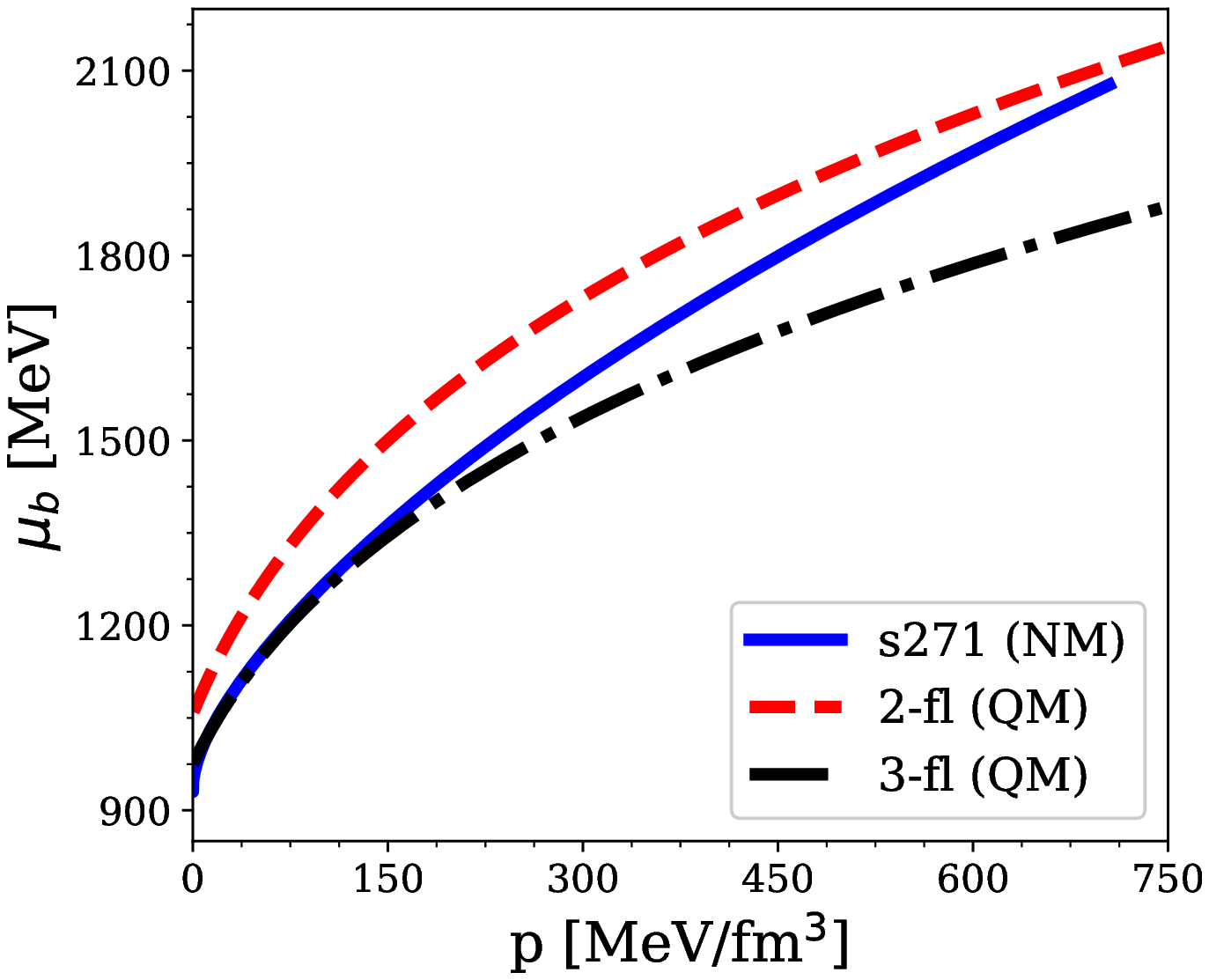}
	\includegraphics[width = 3.2in,height=3.0in]{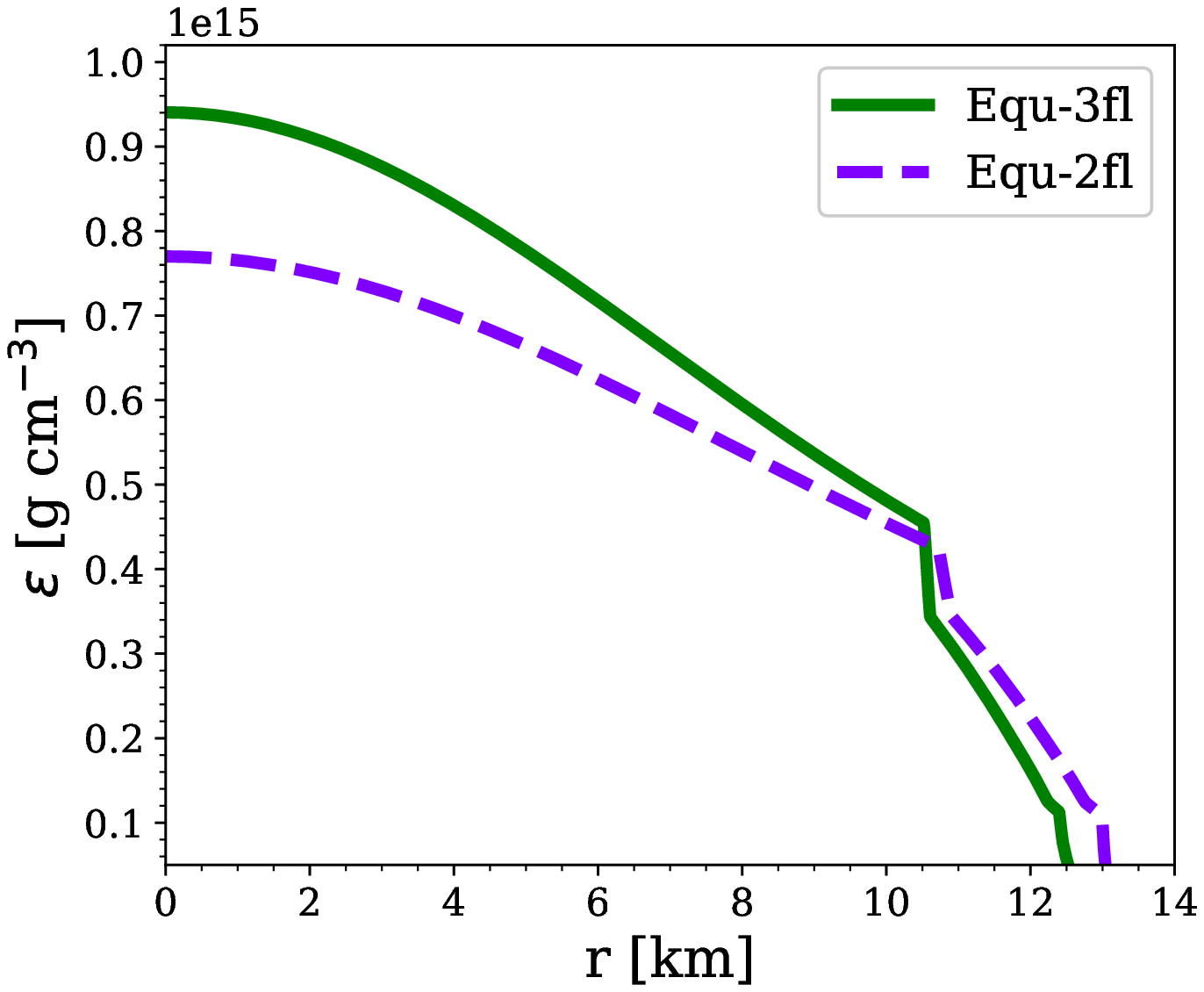}
	%\hskip .4 cm
	%\includegraphics[width = 3.2in,height=2.5in]{den-rad.eps}
	%
	%\hspace{0.5cm} \scriptsize{(a)} \hspace{8.5 cm} \scriptsize{(b)} 
	\caption{a) $\mu_b$ against $p$ curve for NM, 2-f QM, and 3-f QM is shown. The 3-f QM is more stable than NM at high densities. Although, NM is more stable than 2-f matter even at high densities, however, deconfinement would first happen and then the deconfined 2-f matter would settle into 3-f matter eventually. 
		b) The radial Energy density profile inside the star for both 2-f and 3-f matter along the equatorial direction is shown. The profile is obtained by solving the slow rotating star code.}
	\label{den-prof}
\end{figure}

\begin{figure}
	%\vskip 0.2in
	\centering
	\includegraphics[width = 10cm]{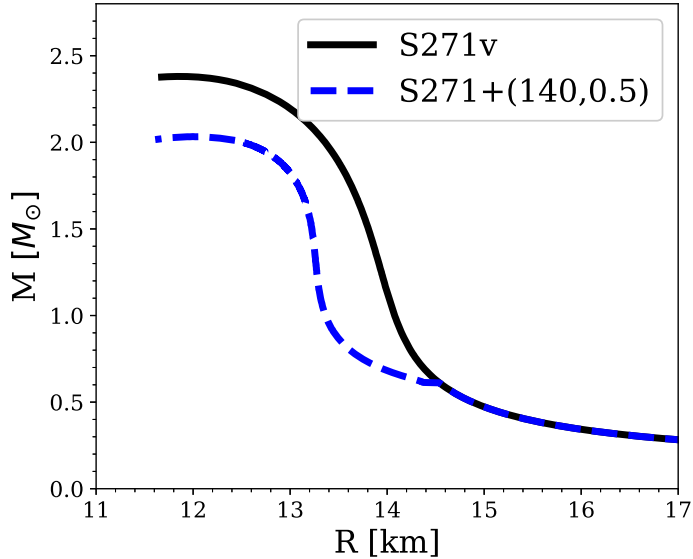}
	\caption {{The mass-radius relations for S271v2 hadronic model is shown with the dotted curve and the mass-radius relation for the HS (S271v2 + 3f Quark matter ($B^{1/4}=140$, $a_4=0.5$)) is demonstrated by the solid curve.} }
	\label{mr}
\end{figure}

The hypothesis of absolutely stable quark matter (SQM) states that 2-f QM is unstable (compared to NM) at low temperatures and densities. However, the extra degrees of freedom in 3-f QM lowers its free energy per quark to the extent that is it could be stable than hadronic matter and 2-f QM (fig \ref{den-prof}(a)) \citep{witten,ouyed}. Such conversions would require the hadronic matter to be heated up or compressed to the point that it deconfines into 2-f QM and eventually decay into 3-f QM. For this, the system must be injected with sufficient energy so that it is energetically favorable for the strong interaction to deconfine hadronic matter into 2-f QM. Such a scenario exists in NS cores, where a sudden abrupt deconfinement of HM in a narrow central region into 2-f QM results in a shock. These propagating shocks provide necessary compressions for 2-f matter formation in the region surpassed by it, thus converting a large chunk of HM to 2-f matter within the NS. The 2-f QM eventually decays to 3-f QM, releasing energy explosively \citep{ouyed,mallick}. Thus the shock-induced combustion in NS is a two-step process. The initiation of such a process involves a density fluctuation at the center of the star, which can arise during mass accretion or glitch in a cold neutron star that is trying to settle into a stable configuration. The shock-induced PT is also likely to occur in BNSM after the first merging during the hypermassive star stage. The deconfinement conversion is a fast process, and it has been previously shown that the speed of such a shock front is close to the speed of light \citep{prasad,prasad1}. The outcome of the aftermath of such shock combustion is that the matter at the core of the star is unstable 2-f QM (as can be seen from fig \ref{den-prof}(a)), and it converts to 3-f QM by conversion of excess down quarks to strange quarks to attain absolute stability. The conversion of 2-f to 3-f matter proceeds as slow combustion where weak interaction governs the combustion process. Once beta equilibrium is attained for a particular radial point, the combustion front proceeds forward. The NM to QM conversion happens till the point where NM is more stable than QM (the critical point (CP)). The crossing point of NM and the 3-f matter is the CP point (shown if fig \ref{den-prof} (a)), and for a given value of $p$, the matter which has lower $\mu_b$ is more stable. Once CP is identified, solving the stellar structure equation gives the corresponding CP inside the star, shown in fig \ref{den-prof} (b). We have used S271 \citep{lala,horo} parameter setting to describe the NM and MIT bag model having quark interaction \citep{chodos,alford,weissenborn} for QM. Both the euqation of state (EoS) is consistent with the recent nuclear and astrophysical constraints. For the QM, the 2-f matter has only up quark with mass ($2$ MeV) and down quark with mass $5$ MeV, whereas the 3-f matter also has strange quarks with mass $95$ MeV. The bag constant is assumed to be $B^{1/4}=140$ MeV, and the strength of quark coupling is $a_4=0.5$.  We plot the mass-radius relation for our choice of EoS in fig \ref{mr} and found the maximum mass of HS, which is $2.03 M_\odot$. It is clear from the fig \ref{mr} that the maximum mass of HS and radius of a $1.4 M_\odot$ HS ($R_{1.4}=13.25$ km) are in agreement with the estimated limit set by the observation of GW170817 NS-NS merger \citep{fattoyev,rezzolla18, zhou}.

We perform our calculation for a rotating axisymmetric star. In our calculation, we have used Hartle's slow rotation approach for constructing the rotating star \citep{hartle1,hartle2}. This approach is quite accurate for stars whose rotational speed is up to some hundreds of Hz. We have constructed stars with a rotational speed of $50$ and $500$ Hz in our calculation. The NM to 2-f matter scenario is discussed previously by us \citep{prasad,prasad1}, where the fast PT front following the shock deconfine the NM to 2-f QM. It leaves a unique signal in the form of a gravitational wave of strain $10^{-21}-10^{-22}$ and frequency of a few hundreds of kHz \citep{prasad1}.

The beta equilibrated 3-f matter is charge neutral and the baryon conservation still holds. This can be written as,

\begin{gather}
2n_{u} = n_{d}+n_{s}+3n_{e^-} \\ 
3n_b =n_{d}+n_{s}+n_{u}
\label{brn-cons}
\end{gather}

where $n_b$, $n_{u}$, $n_{d}$, $n_{s}$ and $n_{e^-}$ are baryon, up quark, down quark, strange quark and electron number densities respectively.

 For the quark EoS, we follow the modified MIT Bag model expressed by the Grand potential \citep{alford,weissenborn}

\begin{equation}
	\Omega_{QM}=\sum_f \Omega_f +\frac{\mu^4_b}{4\pi^2}(1-a_4)+B, \\
\end{equation}
where $\Omega_f$ is  the grand potentials of non-interacting free quark and electron gas and is given by
\begin{equation}
\Omega_f=-\frac{1}{4\pi^2}\Bigg[\mu_f\sqrt{\mu^2_f-m^2_f}\Big(\mu^2_f-\frac{5 m^2_f}{2}\Big)+\frac{3}{2}m^4_f\ln{\frac{\mu_f\sqrt{\mu^2_f-m^2_f}}{m_f}}\Bigg].\\
\end{equation}

The number density then becomes 
\begin{equation}
n_f=-\frac{\partial\Omega_f}{\partial\mu_f}=\frac{1}{\pi^2}(\mu^2_f-m^2_f)^{3/2}-\frac{\mu_b^3}{27\pi^2}(1-a_4),
\label{grand-po}\end{equation}

where the index $f$ stands for $u$, $d$, $s$ quarks and electron. $a_4$ is the strong interaction correction parameter, and $B$ is the bag parameter of the model. $\mu_b$ is baryon chemical potential which is the sum of u,d, and s quarks chemical potential ($\mu_u+\mu_d+\mu_s$). From the eqn. \ref{grand-po} we find the number density of quarks in terms of their chemical potential.

We consider a frame of reference in which the conversion front is stationary. It is also assumed that the volume in which conversion takes place to be much smaller than the total quark matter region. Therefore, the problem can be treated in one dimension. The density fluctuation deconfines the nuclear matter at the core of the star.
Once the shock wave has propagated out, we have 2-f QM at the star core and the NM near the surface.
The quark-nuclear transition point depends on the EoS, and for our choice of it is around $10$ km. The unstable 2-f core has an excess of down quarks, which convert into the strange quarks via weak decay as long as $\mu_k=\mu_d -\mu_s$ is non zero, the decay equations are given by

\begin{gather}
d\rightarrow u + e^- + \nu_{e^-},\\
s\rightarrow u + e^- + \nu_{e^-,}\\
d+u\quad\rightarrow\quad s+u.
\label{wk-decay}
\end{gather}

The conversion stops when $\mu_d=\mu_s$, and we have stable 3-f quark matter. At equilibrium, the down and strange quark's chemical potential can be written in terms of the chemical potential of up quark and electron, assuming that neutrinos will escape freely from the star \citep{abhijit}. Therefore, we have
\begin{equation}\label{ch-pot-equ}
\mu_{d}=\mu_{s}=\mu_{u}+\mu_{e^-}.
\end{equation}

The conversion process starts at the center of the star and goes towards the region's surface, filled with the unstable 2-f matter. In the rest frame of the combustion front, the 2-f QM is at the right-hand side (RHS), and the equilibrated 3-f QM is on the left.
To parametrize the PT, we define a quantity `$a$' as function of radial co-ordinate $r$ taking center of star as the origin,
\begin{equation}\label{a}
a(r)=\frac{n^{2f}_{k} (r)-n^{3f}_{k} (r)}{2n_b(r)},
\end{equation}
 where $n_k=\frac{1}{2} (n_d-n_s)$, superscript 2f and 3f denotes the  2-f and 3-f matter respectively. The region which has only 2-f matter, strangeness is zero and therefore $n_k=\frac{n^{2f}_d}{2}=n^{2f}$ (neglecting electron chemical potential). 
	Very close to the conversion front (however, in the 3-f region), the 3-f matter is just generated and is still not in beta equilibrium, $n_k$ takes a value given by $n_k=n^*_k $. Far from the front in the 3-f region, we get a value for $n_k$, which is asymptotically equal to $n^{3f}_k$.

The conversion process can be studied through two processes: first is the decay of down quark into strange quark via weak decay (eqn \ref{wk-decay}), and the second diffusion of strange quark into the matter. Defining $n_b R$ as the decay rate, the weak decay can be mathematically written as \citep{olinto}

$$ n_b R=-\frac{d n_{d}}{d t}=\frac{d n_{s}}{d t},$$
$$\Rightarrow 2 n_b R=-\frac{d(n_{d}-n_{s})}{d t},$$
$$\Rightarrow R=-\frac{d}{d t}\Big(\frac{n_{d}-n_{s}}{2n_b}\Big),$$
which finally gives
\begin{equation}\label{dcy-eq}
\frac{d a}{d t}=-R(a).
\end{equation}
the weak decay equation.

The equation for diffusion of strange quark is given by
\begin{equation}\label{diff-eq}
\frac{d a}{d t}=D\frac{d^2 a}{d r^2}.
\end{equation}
where D is diffusion coefficient \citep{olinto}. D is expressed as
\begin{equation}
	D\simeq 10^{-3}\Bigg(\frac{\mu_b}{T}\Bigg)^{2}\: cm^2 s^{-1},\label{D}
\end{equation}
with the baryon chemical potential $\mu_b$ and temperature $T$ are in MeV.
Therefore, the change in $a$ with time is given by (using \ref{dcy-eq} and \ref{diff-eq})
\begin{equation}
\frac{d a}{d t}=D\frac{d^2 a}{d r^2}-R(a).
\label{rate-dif}
\end{equation}

From the usual definition of total derivative, we have
\begin{equation}
\frac{d a}{d t}=\frac{\partial a}{\partial t}+(\vec{v}\cdot\vec{\nabla}a),
\end{equation}
and for one dimensional steady flow $\frac{\partial a}{\partial t}=0$, then this reduces to
\begin{equation}\label{adot}
\frac{d a}{d t}=v \frac{d a}{d r}.
\end{equation}
Therefore, using  eqn \ref{rate-dif} and \ref{adot}, we finally have
$$D\frac{d^2 a}{d r^2}-v\frac{d a}{d r}-R(a)=0.$$

This can be simply denoted by
\begin{equation}\label{fnl- eq-x}
D a''-va'-R(a)=0\quad\text{where}\quad a''\equiv\frac{d^2 a}{d r^2};\quad a'\equiv\frac{d a}{d r}.
\end{equation}

{\it The Boundary Conditions}

Let us assume that at conversion front is at a distance of $r=\bar{r}$ from the center. In the rest frame of the front, the 2-f matter will come from the right, get converted into 3-f matter, and goes towards the left. We write the equation of conservation of flux current at $\bar{r}$ as
$$ n_{3f}v_{3f}=n_{2f}v_{2f}.$$
Eqn \ref{fnl- eq-x} can be integrated over small volume of cylinder whose axis coincides with $r$ to give
$$ \int_{V} a'' dV-\int_{V}v a' dV-\int_{V}R(a)dV=0.$$ 
Now assuming length of cylinder is $2\epsilon$ and $\epsilon\rightarrow 0$, the above equation can be written as
$$\lim\limits_{\epsilon\rightarrow 0} \int_{\bar{r}-\epsilon}^{\bar{r}+\epsilon}D a'' dr-\lim\limits_{\epsilon\rightarrow 0} \int_{\bar{r}-\epsilon}^{\bar{r}+\epsilon}v a' dr-\lim\limits_{\epsilon\rightarrow 0} \int_{\bar{r}-\epsilon}^{\bar{r}+\epsilon} R(a) dr=0.$$

In the limit of vanishing $\epsilon$ the third term in the above equation goes to zero. The first two can becomes 
$$\lim\limits_{\epsilon\rightarrow 0}\Big\{D a'\Big |^{\bar{r}+\epsilon}_{\bar{r}-\epsilon}-va\Big |^{\bar{r}+\epsilon}_{\bar{r}-\epsilon} \Big\} =0.$$ This is explicitly written as 
$$\quad D\lim\limits_{\epsilon\rightarrow 0}[a'(\bar{r}+\epsilon)-a'(\bar{r}-\epsilon)]-v \lim\limits_{\epsilon\rightarrow 0}[a(\bar{r}+\epsilon)-a(\bar{r}-\epsilon)]=0, $$
or $$D [a'|_{r}-a'|_{l}]=v[a|_{r}-a|_{l}].$$
From our choice of reference frame, the right hand side has only 2-f matter, therefore, we have 
$$a\big|_{r}=\frac{n^{2f}}{n^{3f}}=a_r$$ and $$a'\big|_{r}=0.$$ In the left hand side there is 3-fl matter then $a'|_{l}=a'(\bar{r})$. With the same argument at $\bar{r}=0=r$, we have

$$a\big|_l=\frac{n^*_k}{n^{3f}}=a^*=a_l$$
\begin{equation}\label{bc-2}
a'(0)=-\frac{v}{D}(a_r-a_l).
\end{equation}
 At the conversion front $a(r)$ becomes equal to $a^*$ and as the matter is still not in beta equilibrated $\mu_k$ is non-zero. However, its gets equilibrated very quickly as the front velocity is assumed to be fast enough. The value of $a^*$ which occurs very close to the conversion front  is given by
	$a^{*}=\sqrt{\dfrac{2\Delta\mu_b\chi^{Q}_k}{n^{3f}}}$ \citep{alford15}. $\Delta\mu_b $ is the difference in the chemical potential across the front and $\chi^{Q}_k=\frac{\partial n_k}{\partial \mu_k}$ is the susceptibility. In the left of the front (3-f region) $n_d\approx n_s$ which means $a_l\approx 0$ whereas on the right (2-f region) $n_{2f}\approx n_{3f}$ meaning $a_r\approx1$. From the above discussion one sees that  $a'<0$, and $a$ can take values only between $0$ and $1$, which means that $a$ is a monotonic decreasing function. 
	
	Coming back to the main differential equation (DE), we can write  the decay rate $R(a)$ at a temperature $T$ as \citep{alford15,drago15}.

\begin{equation}\label{R}
R(a)\simeq\frac{128}{27 \times 15\pi^3} G^2_F\cos^2\theta_c\sin^2\theta_c\mu^5 (a^3+\zeta a)=\frac{a^3+\zeta a}{\tau}.
\end{equation} 
where 
\begin{equation}
\zeta=\frac{9 \pi^2 T^2}{\mu^2}
\label{zeta}
\end{equation}
\begin{equation}
\tau=\Big[\frac{128}{27 \times 15\pi^3} G^2_F\cos^2\theta_c\sin^2\theta_c\mu^5\Big]^{-1}\simeq 1.3\times 10^{-9}\Big[\frac{300\quad MeV}{\mu}\Big]^5 sec.
\label{tau}
\end{equation}

Redefining $x$ as $x=\eta \xi$ and $\eta=\frac{D}{v}$, DE \ref{fnl- eq-x} is written as
\begin{equation}\label{fnl-eq-xi}
\frac{d^2 a}{d\xi^2}-\frac{d a}{d\xi}-g (a^3+\zeta a)=0.
\end{equation}
where 
\begin{equation}\label{g}
g\equiv\frac{D}{\tau v^2}.
\end{equation}
The BC (\ref{bc-2}) then takes the form  
\begin{equation}\label{bc2-1}
\frac{d a}{d\xi}(0)\equiv\frac{d a}{d \xi}\Big|_0=-(a_r-a_l).
\end{equation}

\begin{figure}
	\centering
	\includegraphics[width = 10cm]{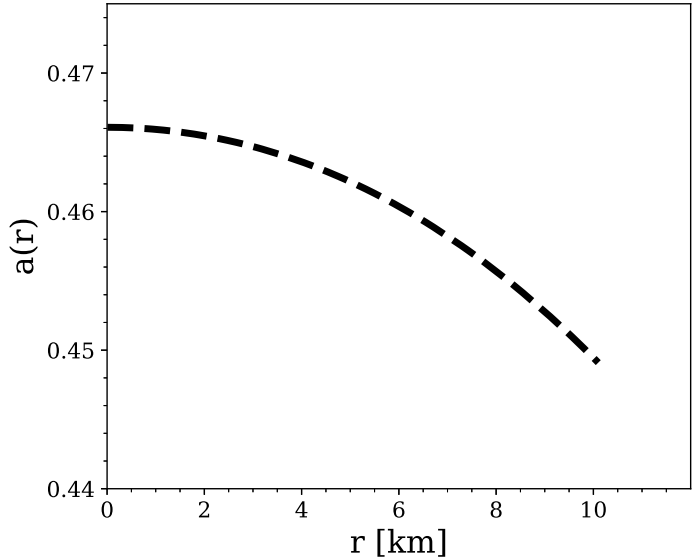} 
	\caption{The ratio $a(r)$ inside star at a distance $r$ from the center of the star.}
	\label{avr}
\end{figure}

 The DE \ref{fnl-eq-xi} can be solved numerically using BC \ref{bc2-1} for a given value of $a_r$, $a_l$ and $g$. We solved the TOV equation for slowly rotating star to get density as the function of $r$ of quark star (fig \ref{den-prof} (b)). In the process of conversion the baryonic mass $M_b$ remains constant. In fig \ref{den-prof} (b) we have plotted energy density vs radius for both along the equatorial and polar direction. For the Baryonic mass of the star $M_B=2.08678 M_{\odot}$, the gravitational mass of 2-f and 3-f stars are $M_{2f}= 1.8008 M_{\odot}$ and $M_{3f}= 1.7955 M_{\odot}$ respectively.
After finding number density, we can find $a$ at any radial point and as shown in fig \ref{avr}. The value of $a$ is maximum at the center, and as we proceed outward of the star $a$ decrease and it is minimum at the point where the PT stops, and NM is stable.

\begin{figure}
	\vskip 0.2in\centering
	\includegraphics[width = 10cm]{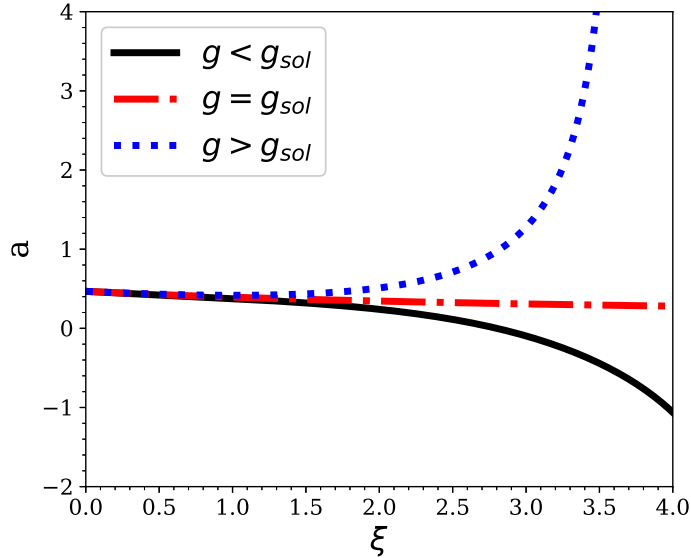}
	\caption{ Numerical solution of DE \ref{fnl-eq-xi} for $a=0.4671435$. In this graph we have plotted $a$ vs $\xi$ for three different values of $g$: $g=1.1<g_{s}$, $g=1.3518=g_{s}$ and $g=1.1>g_{s}$; here $g_{s}$ correspond to value of $g$ for which solution is valid..}
	\label{g_a0}
\end{figure}

 The next task is to calculate the value of $g$. We start with some value of $g$ and then solve the DE \ref{fnl-eq-xi}. To get an unique solution of the DE we must know the nature of $a$ in the star. As discussed earlier $a$ can take value between $0$ and $1$ which is always positive. Therefore, $\frac{da}{d \xi}$ will be negative (from eqn \ref{bc2-1}), and therefore $a$ is a monotonically decreasing function. For given $a$, $a_l$ $a_r$, we solve DE \ref{fnl-eq-xi} for different value of $g$, however, only one of $g$ gives the correct solution $g_{s}$. If $g< g_{s}$ then solution of DE undershoots and if $g>g_{s}$ then solution overshoots (see fig \ref{g_a0} (a)). By narrowing down these value of $g$ we find $g_{s}$ at every radial point. After finding $g_{s}$ we can find conversion velocity $v$ from the eqn \ref{g} and is given by
	\begin{equation}\label{v}
	v=\sqrt{\frac{D}{\tau g_{s}}}.
	\end{equation}
	
	We solve DE \ref{fnl-eq-xi} at every radial point and find value of  $g$ as described above and shown in fig \ref{g_a0} (a). We start with a guess value of g and solve the DE by trial and error method to find the solution of $g$ within error estimates. Once we find $g_s$, we then calculate the velocity of the front.

To find the velocity, one needs to find the number densities of various species in the star. To do that, we make use of the assumption that with no mass loss during the conversion process, and therefore baryon number conservation and charge neutrality still hold (eqn \ref{brn-cons}). 
Using the density profile inside the star and using eqns  \ref{grand-po} and \ref{ch-pot-equ}, we can find the number density of various species inside the star function of $r$. In our calculation, we assume up and down quark are massless but strange quark has mass $m_{s}\approx 100 MeV$, and we also assume that electron number density is negligible.% The equations for finding the number density is as follows 

%\begin{equation}\label{nd-1}
%n_{u}=n_b.
%\end{equation}

%\begin{equation}\label{nd-3}
%n_{d}=\frac{(( n_{s}\, \pi^2)^{\frac{2}{3}}+m_s^2}{\pi^2}.
%\end{equation}
%\begin{equation}\label{nd-4}
%n_{s}+\frac{[(n_{s}\, \pi^2)^{\frac{2}{3}}-m_{s}^2]^{\frac{3}{2}}}{\pi^2}=2\, n_b.
%\end{equation}

\begin{figure}
	%\vskip 0.2in
	\centering
	\includegraphics[width = 3.2in,height=2.5in]{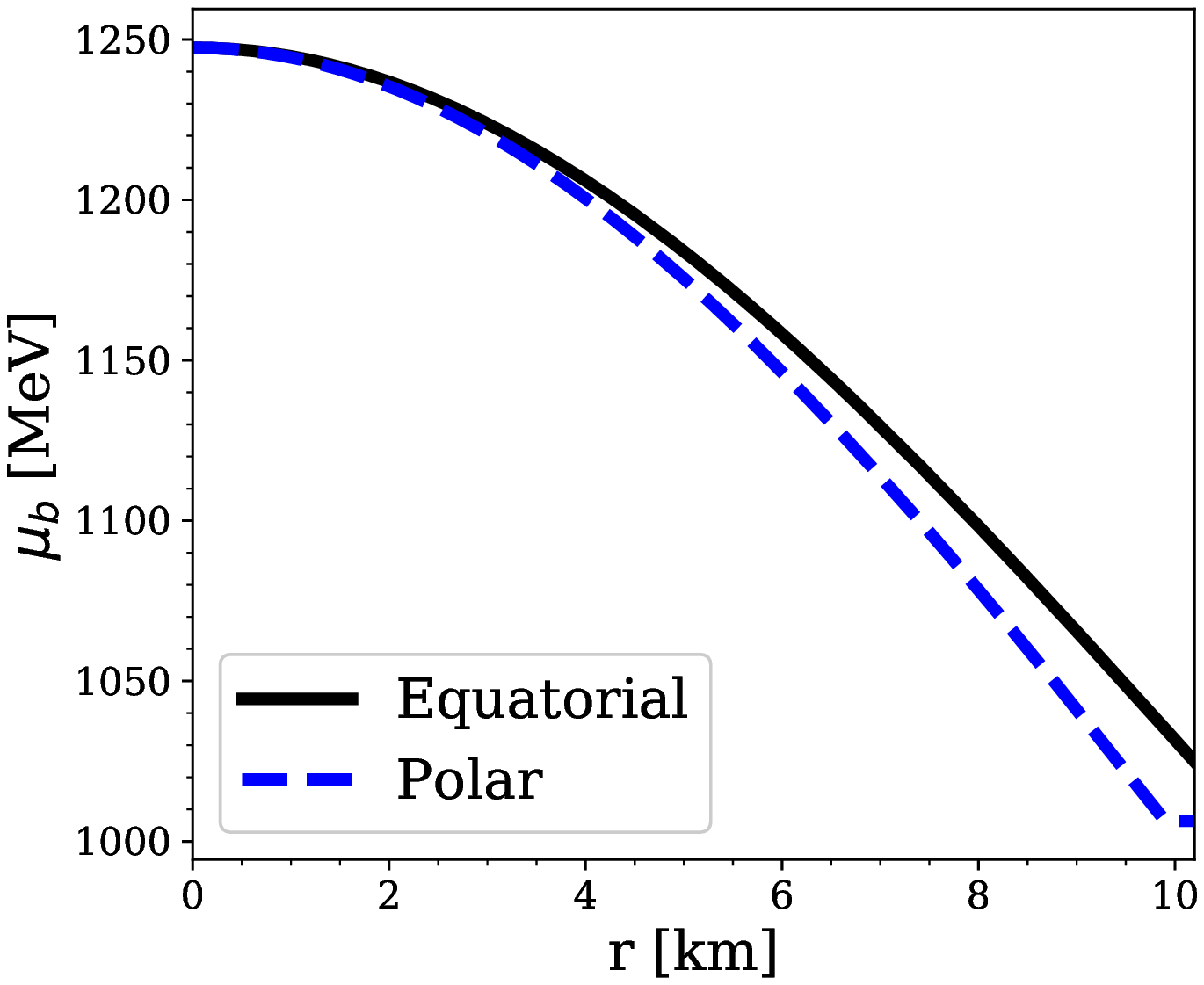}
	%\hskip .4 cm
	\includegraphics[width = 3.2in,height=2.5in]{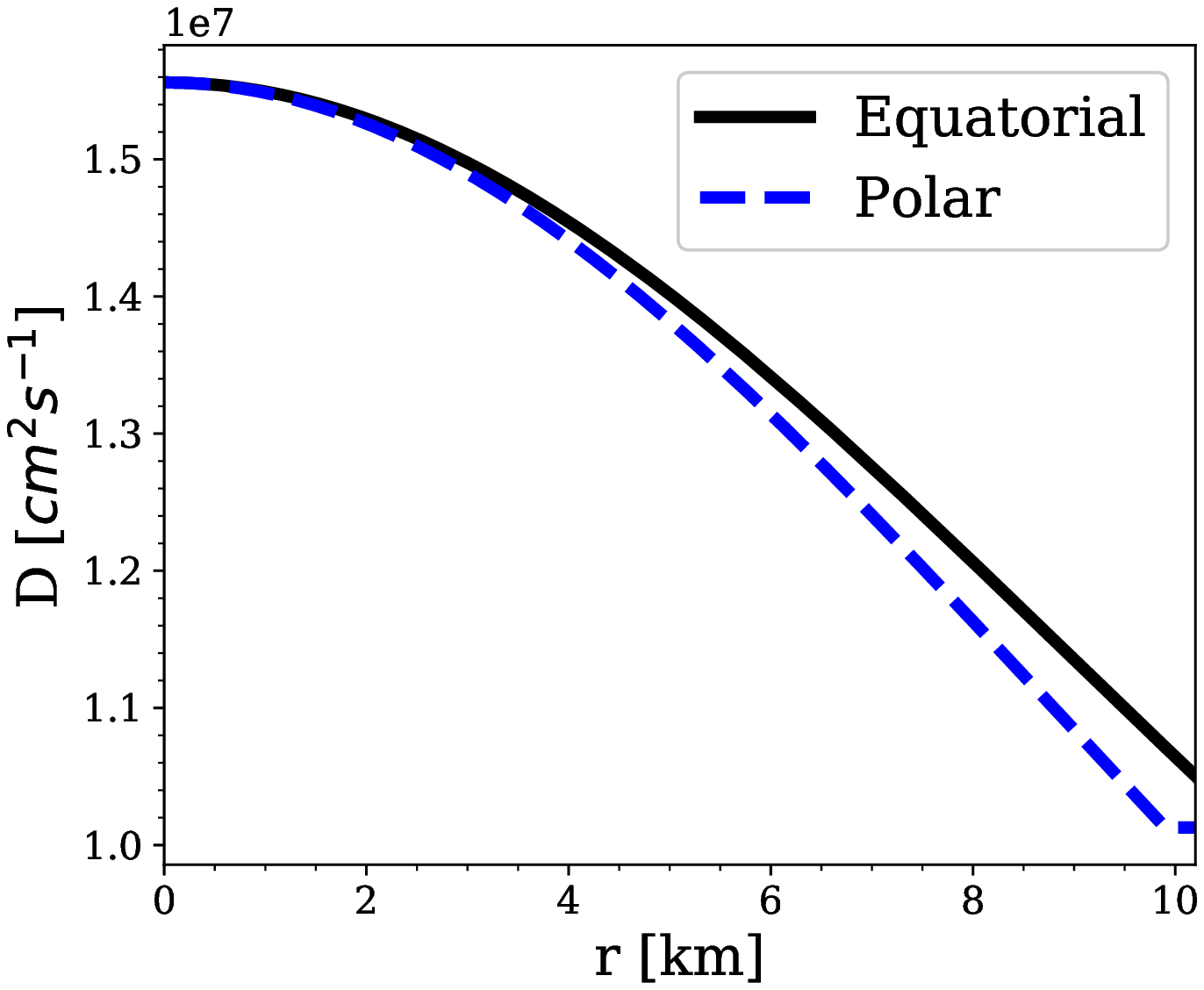}
	
	\hspace{0.5cm} \scriptsize{(a)} \hspace{8.5 cm} \scriptsize{(b)}
	\caption{a) Baryonic chemical potential $\mu_b$ as function of r inside the star along the equatorial and polar directions is illustrated. b) Diffusion coefficient $D$ as function of r is shown inside the star along the equatorial and polar directions.}
	\label{D-r}
\end{figure}

\begin{figure}
	\centering
	\includegraphics[width = 10cm]{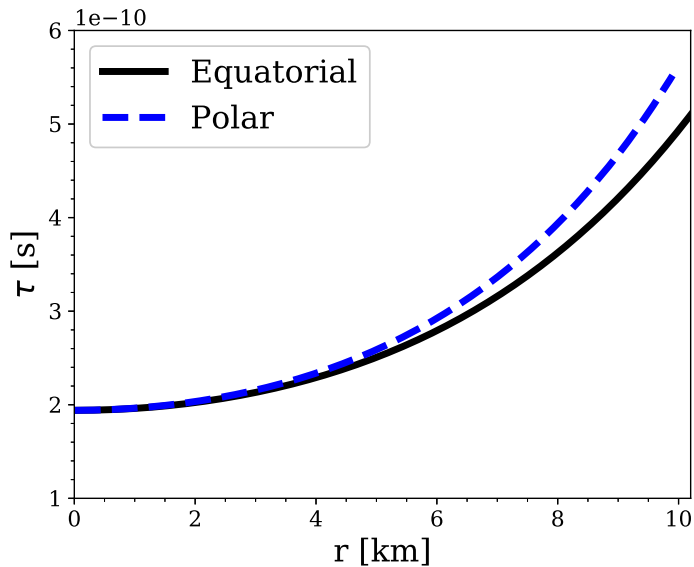} 
	\caption{Variation of $\tau$ inside the star at a distance $r$ from the center of the star.}
	\label{fig-tau}
\end{figure}

For the calculation of the velocity, along with $g_s$, two other quantities are also needed to be calculated, D and $\tau$. As is shown in eqn \ref{D}, D is a function of $\mu_b$, and from the star profile, we find $\mu_b$ as a function of $r$. Therefore, we can get $D$ as a function of $r$. The variation of $\mu_b$ as a function of $r$ is shown in fig \ref{D-r} (a). As expected, the chemical potential falls off as we go outwards of the star. The fall is smooth and monotonic till the point where PT stops. Beyond the PT point, $\mu_b$ gets saturated around $1000$ MeV. D also depends on the star temperature, and for a cold NS, we assume that the star's temperature is constant throughout and assumed its value to be equal to $10^{-2}$ MeV.

Therefore, with the assumption that D directly depends on $\mu_b$, D's behavior closely resembles that of $\mu_b$ as shown in fig \ref{D-r} (b). It starts with a high central value at the star center and falls off as we proceed outwards. Again, there is a discontinuity at the PT point inside the star.
The final parameter that remains to be calculated is $\tau$. The value of $\tau$ is inversely proportional to $\mu_u^5$, as given in eqn \ref{tau}. The up quark chemical potential also depends on the radial distance from the star center. Calculating the up quark chemical potential, the variation of $\tau$ as a function of radial distance is shown in fig \ref{fig-tau}.

\begin{figure}
	%\vskip 0.2in
	\centering
	\includegraphics[width = 3.2in,height=2.5in]{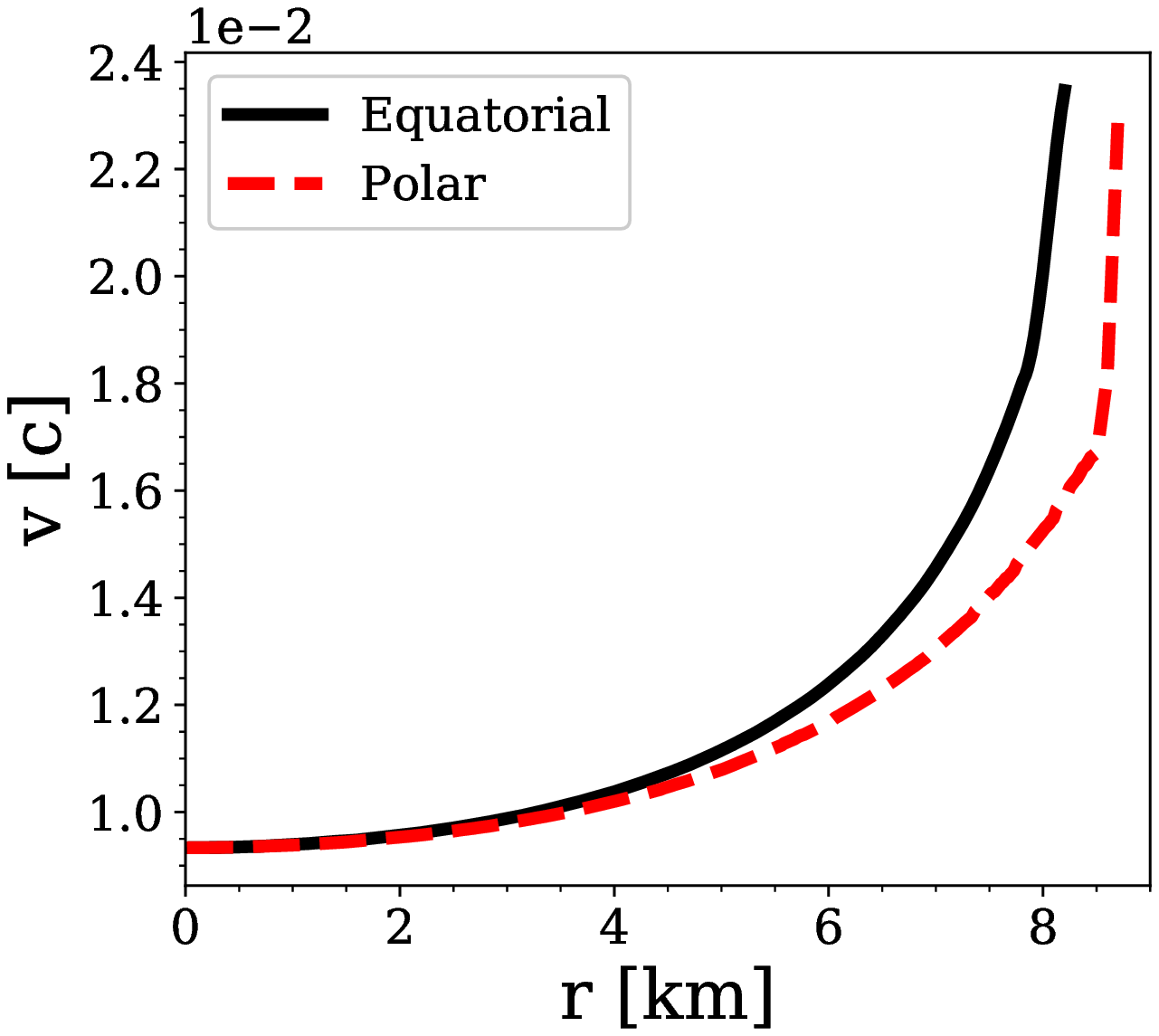}
	%\hskip .4 cm
	\includegraphics[width = 3.2in,height=2.7in]{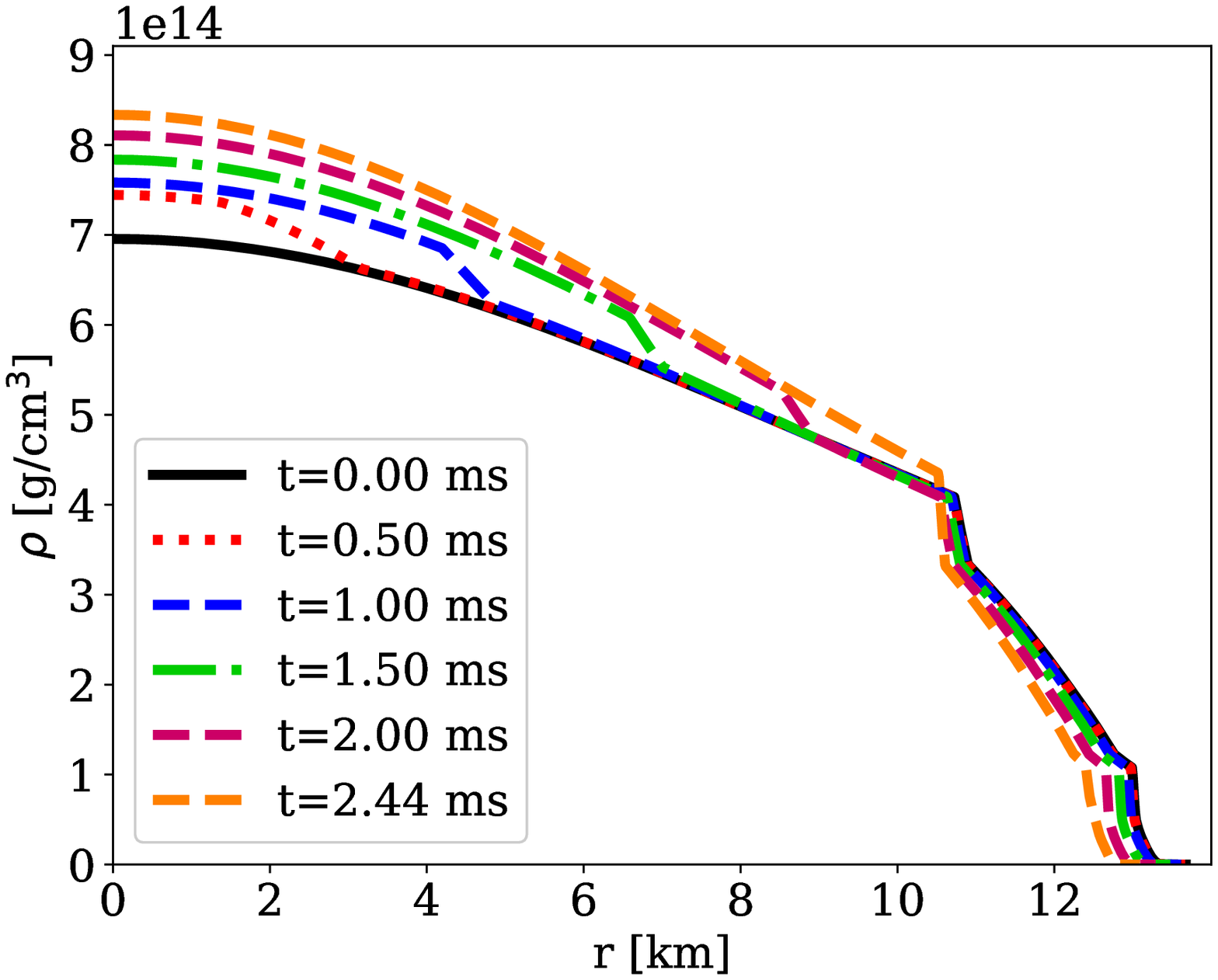}
	
	\hspace{0.5cm} \scriptsize{(a)} \hspace{8.5 cm} \scriptsize{(b)}
	\caption{a) Conversion velocity inside star at a distance $r$ from the center of the star in equatorial and polar direction. b) Density profile along equatorial direction inside the star at various time. The conversion front can be seen as a discontinuity.}
	\label{front}
\end{figure}

\section{Gravitational waves}

Having calculated all the necessary quantities, we can now calculate the velocity of the conversion front from the center to the star's surface. The conversion front velocity as a function of radial distance from the star center is presented in figure \ref{front} (a). The value of $v$ at the star center is about $9 \times 10^{-3}$ c, and initially, as we move towards the surface of the star, it monotonically increases up to CP; after that, it shoots up to a value $2.3 \times 10^{-2}$ c. Knowing the velocity profile, we can find the position of conversion front and total conversion time by integrating $ dt=\frac{d r}{v(r)}$. The total conversion time comes out to be $2.44$ ms. As the front velocity depends on the radial position, we can find the front position at any given time from equation $v(r)=\frac{d r}{d t}$. We assume that at time $t=0$, the 3-f seed originates at the center of the star. Knowing the velocity, we locate the conversion front after a short time interval ($dt=10^{-5}$ sec). At this position, the right side of the front has 2-f matter, and the left side has 3-f matter. 
Using the combustion front location and ensuring that the star's baryonic mass remains constant, we construct an intermediate star. The star has a burnt central 3-f QM region (till the front location), an unburnt 2-f QM (unstable) non-central region till the CP, and beyond it, we have an HM outer region.  
Thus, we get the star's density profile for a particular time, and integrating it gives the mass quadrupole moment (QMoM). The steps are repeated (till the CP, where finally we only have $\text{3-f} + NM$) to obtain the change in QMoM and the corresponding GW amplitude. The change in density profile as a function of time is shown in figure \ref{front} (b). The density profile is shown for five-time sequences. Once the change in density profile as a function of time is known, the gravitational wave amplitude is given by
\citep{prasad1,zwerger,dimmelheimer,abdikamalov}
\begin{equation}
h_{\theta \theta}^{TT} =\frac{1}{8} \sqrt{\frac{15}{\pi}} \sin^{2}\theta \frac{A_{20}^{E2}}{d},
\end{equation}
where $ \theta$ is the angle between the symmetry axis and the observer's line of sight. $A_{20}^{E2}$ is described as 
\begin{equation}
A_{20}^{E2}= \frac{d^{2}}{dt^{2}} \left( k \int \rho \left( \frac{3}{2} z^{2}-\frac{1}{2} \right)r^{4} dr dz \right),
\end{equation}
with 
$z=\cos{\theta}$ and $k=\frac{16 \pi^{3/2}}{\sqrt{15}}$. $A^{E2}_{ 20}$ is the second time derivative of the mass QMoM of the source and $d$ the source's distance. These formulae have been employed to study the supernova collapse and the microcollapse resulting from the aftermath of PT  for an axisymmetric star \citep{dimmelheimer,abdikamalov}. The evaluation of the double derivative of the quadrupole moment is sensitive to the choice of the time interval of $dt$ and can result in numerical noise. We use the Savitzky \& Golay \citep{savitzky}  filter and carry out the differentiation and smoothing of the data, with an appropriate window size such that it diminishes variations in quadrupole moment data in timescales of $10^{-5}$ s, and differentiation is carried out for variations which are above $10^{-4}$ s. This smoothing of variation suppresses the numerical noise in the GW strain.

\begin{figure}
	%\vskip 0.2in
	\centering
	\includegraphics[width = 3.2in,height=2.5in]{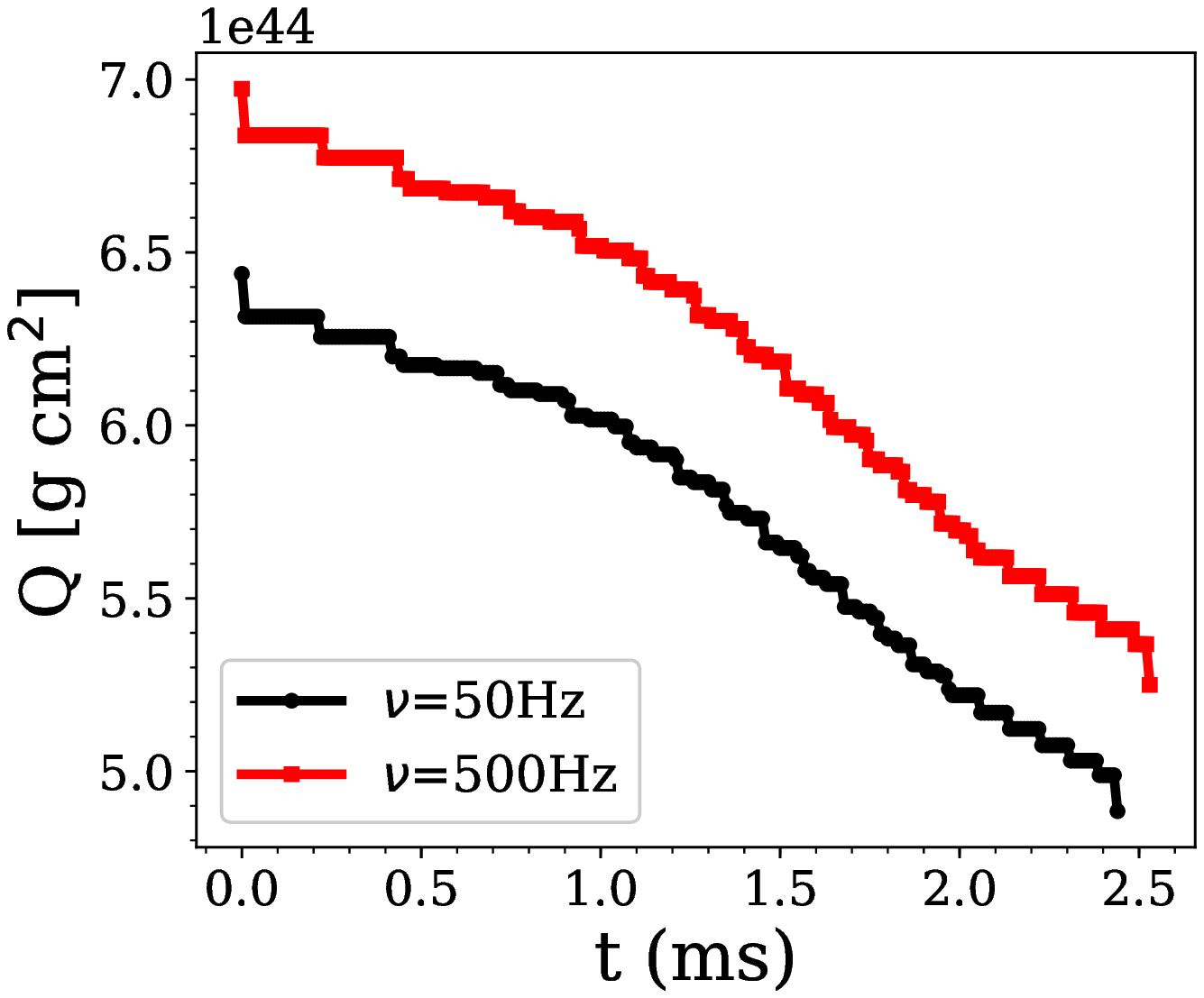}
	%\hskip .4 cm
	\includegraphics[width = 3.2in,height=2.5in]{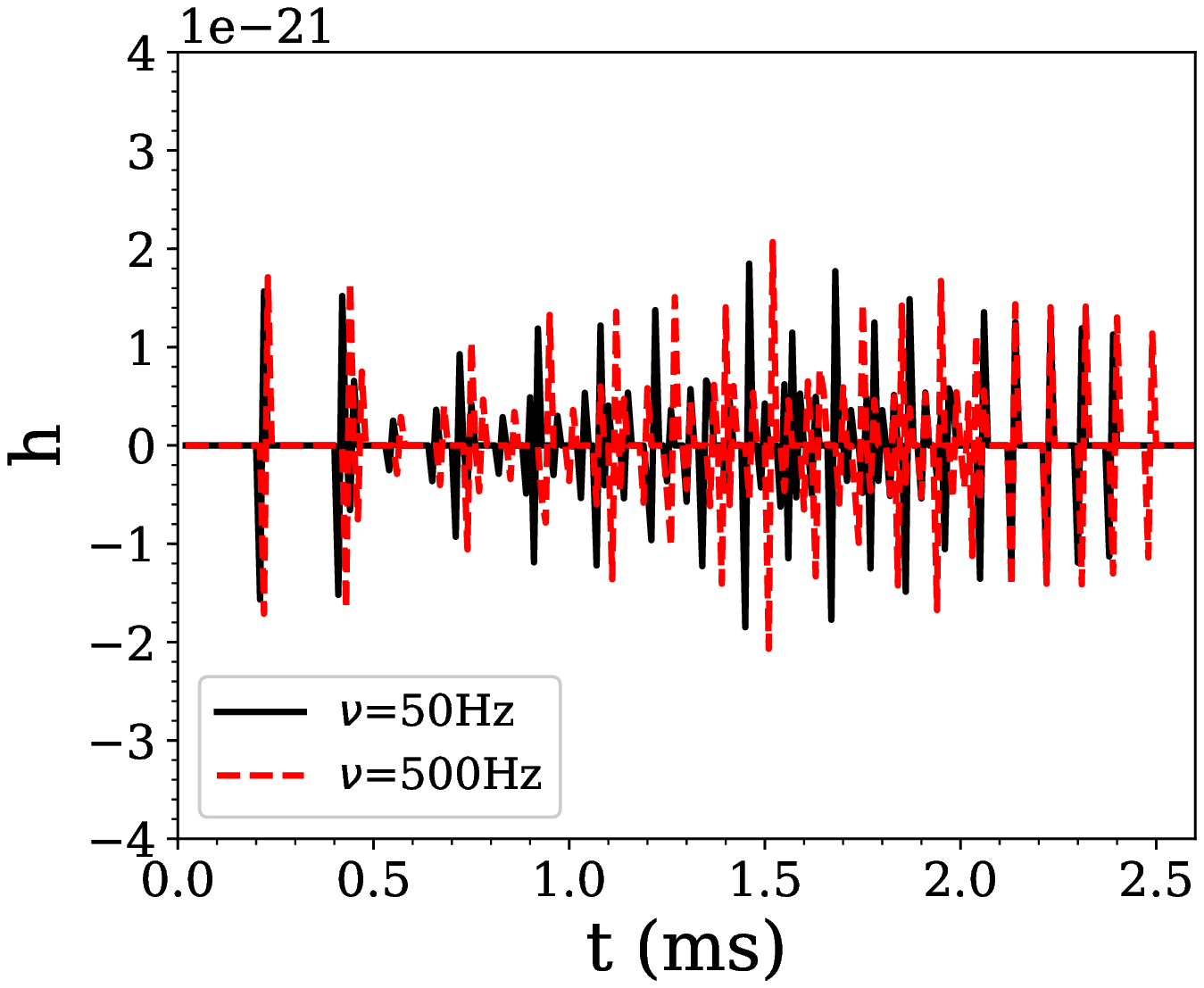}
	
	\hspace{0.5cm} \scriptsize{(a)} \hspace{8.5 cm} \scriptsize{(b)}
	\caption{a) The change in quadrupole moment as a function of time is shown b) The GW amplitude is illustrated for the conversion of 2-f to 3-f matter in HS. The singal is shown for the duration $t=20$ ms to $t=50$ ms.}
	\label{amp}
\end{figure}

\begin{figure}
	\centering
	\includegraphics[width = 10cm]{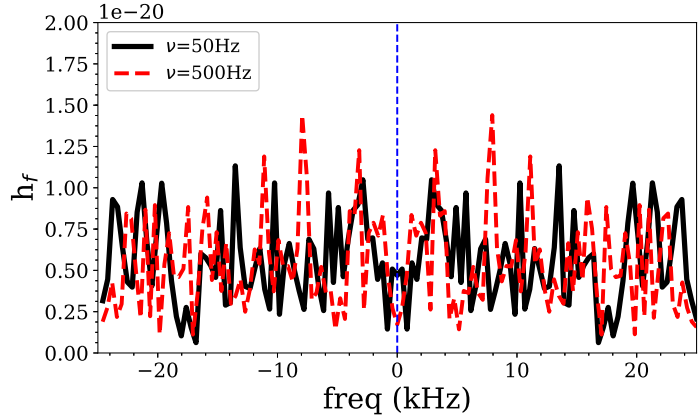} 
	\caption{The ampitude spectral density (ASD) of the GW signal for 2-f to 3-f phase transition is shown for a 1.8 solar mass 2-f star at 1 Mpc distance for $\nu= 50$ Hz and   $\nu= 500$ Hz.}
	\label{asd}
\end{figure}

In fig \ref{amp} (a), we plot the QMoM as a function of time. The quadrupole change for both the stars (with frequency $\Omega=50$ Hz and $500$ Hz) is shown in the plot as the star's density profile is rearranged at every time step the QMoM of the star changes. However, the QMoM decreases globally with time for both the stars. The initial QMoM of the fast rotating star is higher than the slow rotating star. The star is assumed to be located at $d=1$ Mpc with $\theta=\pi/2$. The change of QMoM with time is reflected in the GW amplitude of both the stars. This is shown in fig \ref{amp} (b). The GW amplitude is of the order of $10^{-21}$ for both the stars. Although the QMoM of the fast rotating star is greater than that of the slow rotating star, the change in the QMoM with time is similar for both stars; therefore, the GW strain is also similar. This also means that the energy released from the PT of NS to QS conversion is very similar for both the stars, and the rotation does not play a very significant role in determining the nature of GW generated from PT of NS. The amplitude spectrum is obtained for 1.8 solar mass stars and is shown in fig \ref{asd}; the peaks indicate power contained at different frequencies, the significant peaks are present in $5-25$ kHz. 

Additional studies are performed on 2-f stars of different masses; for this, the 2-f stars are constructed with 1.6, 1.7, 1.9, and 2.0 solar mass. The results are presented in Table 1, wherein the temperature is taken to be $10^{-2}$ MeV. The central density, baryonic mass ($M_{B}$), gravitational mass ($M_{G}$), and equatorial radius ($R_{e}$) of 2-f and 3-f stars together with the time taken for PT are listed in Table 1. It is evident that central density rises due to 2-f to 3-f conversion, and the final 3-f star formed has its gravitational mass less than the initial 2-f star, suggesting energy emission during the process. The total time of PT is seen to be directly related to the mass of the 2-f star, and a massive 2-f star takes a longer time to get converted to a stable 3-f star. This is because massive stars contain a bigger chunk of 2-f content which is unstable and extends up to a larger radius. The GW amplitude for a source at 1 Mpc is of the order of $10^{-21}$, and signal duration is in milliseconds. The amplitude spectrum of the peaks comes out at 10 kHz. All these results further establish the robustness of the 2f-3f PT process.

\begin{table}
	\centering
	\caption{PT in 2-f star's of different masses}
	\begin{footnotesize}
		\begin{tabular}{@{\extracolsep{2pt}}ccccccccccc}
			\hline 
			\hline
			\multicolumn{4}{c}{2-f star}&  \multicolumn{4}{c}{3-f star} & PT : 2f to 3f process\\ 
			\cline{1-4} \cline{5-8} \cline{9-10}
			$\rho_{c}$& $M_{B}$ & $M_{G}$ &$ R_{e}$ & $\rho_{c}$ &$\nu $&$M_{G}$ & $R_{e}$ & time  \\
			$(10^{14} g/cc)$ &  $(M_{\odot})$ &  $(M_{\odot})$& (km) & $(10^{14} g/cc)$ & (Hz) &  $(M_{\odot})$& (km) & (ms)  \\
			\hline
			6.65  &   1.7778 & 1.6043 & 13.74 & 7.77 & 51.7 & 1.6010 & 13.26 &  2.23 &  \\
			7.10  &   1.8979 & 1.7001 & 13.72 & 8.44 & 52.1 & 1.6959 & 13.19 &  2.28  & \\
			7.70  &   2.0263 & 1.8008 & 13.67 &9.40 & 52.8 & 1.7955 & 13.08 &   2.44 & \\
			8.50  &   2.1563 & 1.9008 & 13.58 &10.86& 53.8 & 1.8940 & 12.91 &   2.87  &\\
			9.70  &   2.2898 & 2.0013 & 13.44  &13.95& 56.3 & 1.9927 & 12.55 &  4.47 & \\
			\hline
		\end{tabular} 
	\end{footnotesize}
\end{table}

Till now, we performed our calculation for a cold NS ($T=0.01$ MeV). However, as the PT occurs, it releases energy and can be calculated using $\Delta E = M^{2f}_{G} - M^{3f}_{G}$, where $M_{G}$ denotes the gravitational mass. For the present case, the total energy released comes out to be $1.045 \times 10^{52} ergs$. Most of this energy will dissipate from the system as neutrino's energy and gravitational waves. Some portion of energy released would heat the quark-matter content of the star. We perform a rough estimate of temperature rise associated with the conversion \citep{drago}. Taking the initial temperature to be $10^{-2}$ MeV and the energy budget to heat the star to be $10$ percent of the total energy released and volume to be a spherical shell of radius $10$ km, then the average temperature change comes out to be $\sim 1$ MeV. Therefore, for comparison, temperature values 0.1 MeV and 1 MeV are assumed for 2-f matter, and PT is analyzed.

\begin{figure}
	\centering
	\includegraphics[width = 10cm]{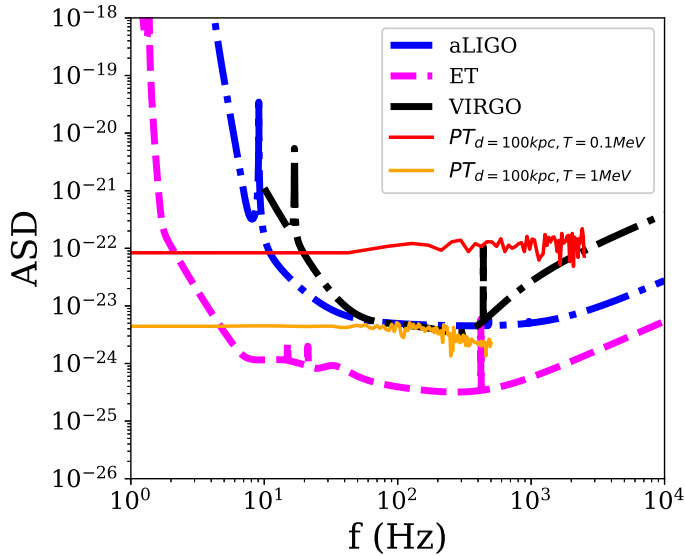} 
	\caption{The ampitude spectral density (ASD) of the GW signal for 2-f to 3-f phase transition is shown for a 1.8 solar mass 2-f star ($\nu= 50 Hz$) at 100 kpc for two different temperature. The noise curve of aLIGO, VIRGO, and Einstein telescope (ET) is shown.}
	\label{asd1}
\end{figure}

The change in the quark EoS for such temperature is minimal, but significant changes arrive due to the dependence of diffusion coefficient and reaction rates on temperature. In Table 2, the results are presented. The PT time strongly depends on the temperature of the 2-f star. PT occurs rapidly for a 2-f star with $T=10^{-2}$ MeV, and its timescale is in milliseconds, whereas for $T=1$ MeV, it takes about one-tenth of a second. The GW wave strain for different source distances is also listed. The GW strain and frequency depend on the temperature of the 2-f star. For a 2-f star having temperature $10^{-1}$ MeV located at 1 Mpc, the GW strain is $10^{-23}$   and the frequency window is in the 10 Hz range, whereas for the $1$ MeV temperature case located at 1 Mpc, the GW strain is $10^{-25}$ and frequency window is in 1 kHz range. The PT in the cooler 2-f star of $10^{-2}$ MeV, the GW is of higher magnitude with the peak frequency located around 10 kHz. The PT in $10^{-1}$ MeV and $1$ MeV 2-f stars are likely to generate detectable GW in the presently operating GW detectors. We obtain the amplitude spectrum density curve of GW signals for source distance 100 Kpcs for two different temperatures and show them along with aLIGO, Virgo, and Einstein Telescope (ET) sensitivity curves (fig \ref{asd1}). The GW signal curve for $10^{-1}$ MeV case lies well above the aLIGO, VIRGO, and ET curves, and with increasing source distance, it will still be in the detection capacity of these detectors. The 1 MeV case can be observed by Einstein Telescope for 100 kpc distance and will be detectable in aLIGO and VIRGO for $d < 100$ kpc. The GW signal duration is directly proportional to the temperature of the 2-f star, whereas the GW signal amplitude is inversely proportional to the temperature of the 2-f star. The faster PT results in a stronger GW signal. Hence GW signals from 2-f stars having $10^{-1}$ MeV will be detectable even for large source distances even beyond $d > 100$ kpc. Among the known radio pulsars, the population of those within our galaxy outnumber the extra-galactic ones \citep{debojoti}. These pulsars are located within a 20 kpc distance. For Crab pulsar ($1.4$ solar mass, $d= 2$ kpc and $\nu = 30 Hz$) the PT time and its associated GW is obtained. Assuming $T=10^{-1}$ MeV the PT time comes out to be $t = 23.8$ ms and the corresponding GW comes out to be $h = 10^{-20}$, and for $T=1$ MeV we get  $t= 0.238$ s and $h=10^{-22}$. This suggests that phase transition happening in the galactic pulsars is likely to be detectable in aLIGO.

\begin{table}
	\centering
	\caption{PT in 1.8 solar mass star with different temperature and distance}
	\begin{footnotesize}
		\begin{tabular}{@{\extracolsep{2pt}}ccccccccccccc}
			\hline 
			\hline
			\multicolumn{3}{c}{2-f star}&  \multicolumn{2}{c}{3-f star} & \multicolumn{6}{c}{PT : 2f to 3f process} \\ 
			\cline{1-3} \cline{4-5} \cline{6-11}
			$\rho_{c}$ & $M_{B}$ & $M_{G}$  & $\rho_{c}$ &$M_{G}$ & Temp & time & $h_{10kpc}$ & $h_{100kpc}$ & $h_{1 Mpc}$ & $h_{f}$ peaks  \\
			$(10^{14} g/cc)$  &  $(M_{\odot})$ &  $(M_{\odot})$ & $(10^{14} g/cc)$ &  $(M_{\odot})$ & MeV &  &  & & &\\
			\hline
			7.70 &   2.0263 & 1.8008  &9.40 &1.7955  & $10^{-2}$ & 2.44 ms &  $10^{-19}$ & $10^{-20}$ & $10^{-21}$ & $10$ kHz\\
			7.70 &   2.0263 & 1.8008 &9.40 &1.7955 &  $10^{-1}$ & 23.4 ms &  $10^{-21}$ & $10^{-22}$ & $10^{-23}$ & 1 kHz\\
			7.70 &   2.0263 & 1.8008 &9.40 & 1.7955 &  $1$ & 0.243 s & $10^{-23}$ & $10^{-24}$ & $10^{-25}$ & 100 Hz \\
			\hline
		\end{tabular} 
	\end{footnotesize}
\end{table}

\begin{figure}
	\includegraphics[width=6in, height=2.5in]{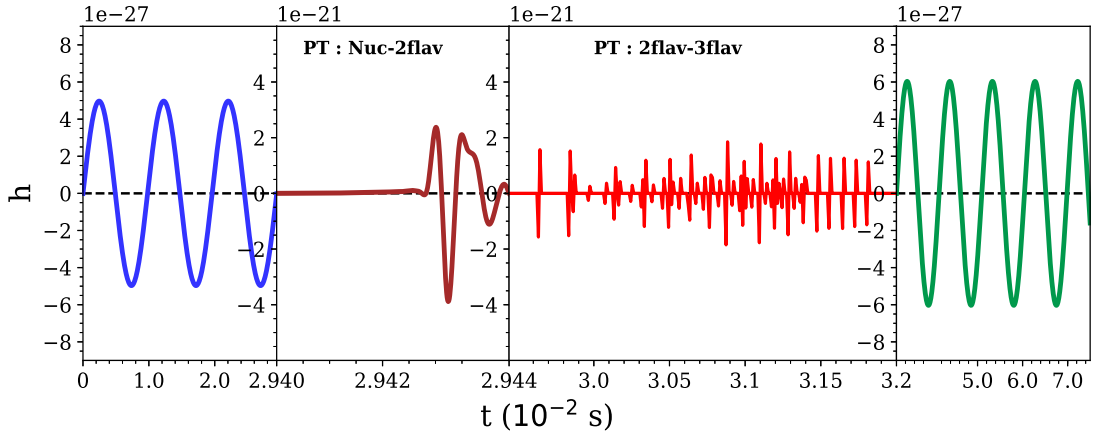}
\caption{We show the strain for the fours stage of combusting NS. The 1st panel shows the GW strain due to NS's rotation ($\nu =51 Hz $). The second panel shows the GW strain of NM to 2-f deconfinement (magnified), leading to an unstable 2-f hybrid star of $\nu =50 Hz$. The third panel shows the conversion of 2-f to 3-f (magnified). Finally, the fourth panel shows the GW from a rotating HS ($\nu =52 Hz$). Note that the strain in different panels is of different strength. The NS, 2-f HS,  and 3-f HS have the same baryonic mass ensured during the PT process, and the rotational frequency is slightly different, although assuming a fixed central angular velocity value due to the difference in the structure of the star in three different phases.}
	\label{tot-wave}
\end{figure}

Usually, pulsars emit continuous low amplitude signals due to their rotation and deformation, usually monochromatic. When PT happens in such a system, two short-lived signals emerge, one from the deconfinement of NM to 2-f QM and others from the 2-f to 3-f QM conversion, and the timescale is in microseconds, and milliseconds respectively. Eventually, the newly formed hybrid star also emits continuous monochromatic GW. The stars formed in the PT process have the same baryonic mass as NS but differ only in gravitational mass and rotation frequency. We give the overall picture of the GW signal coming from an NS, its PT, and the aftermath in fig \ref{tot-wave}. For GW emission, a non-axisymmetric deformation supported by a magnetic field or elastic stress such as the shape is triaxial or else the neutron star's rotation axis should be different from the symmetry axis \citep{ferrari}. The neutron stars studied in the present work are axisymmetric and rotating about their symmetry axis. Here we provide a rough estimate of possible GW strain due to rotation with ellipticity induced by rotation. The ellipticity due to rotational deformation can be estimated with a simple expression, $ \epsilon = \frac{ \omega^{2} R^{3}}{GM}$ which comes out to be $2.2 \times 10^{-3}$ for NS and $1.8 \times 10^{-3}$ for HS. The magnitude of the order of the GW is given by $h_{0}= \frac{4 G}{c^{4} D} I \epsilon \omega^{2}$ \citep{ferrari,poisson}. The GW strain comes out to be $10^{-27}$ for the initial neutron star and final 3-f hybrid star, and this amplitude value is an overestimate due to the high ellipticity values. The actual ellipticity associated with non-axial deformations has been suggested to be in the range $10^{-4}-10^{-7}$ \citep{aasi,lansky}, which would lead to GW order in the range $10^{-28}-10^{-31}$. Detection of such short burst GW signals corresponding to PT and their continuous counterparts would imply that the degree of freedom of matter at high densities are quarks rather than hadrons. It would also imply that NS does undergo a confinement-deconfinement PT at such densities. Such signatures are a unique feature of shock-induced PT, where the combustion front follows a shock. We should mention that we have done the calculation for cold NSs, which undergo sudden density fluctuation due to accretion and glitch. The PT scenario is a more likely scenario at the end of a supernova in the proto-neutron star or hypermassive NS formed after the binary merging of neutron stars. Our future aim is to calculate GW signatures from such scenarios.

\subsection*{Summary and Conclusion}
In this work, we have studied the PT of NM to stable 3-f QM, and especially we have calculated the template of GW amplitude that can be emitted from the conversion of the 2-f to the 3-f matter at the star core. The weak reaction converts 2-f matter to 3-f matter, which follows a shock-induced deconfinement of NM to 2-f matter. The shock-induced PT happens due to a sudden density fluctuation at the star core. As the EoS of 2-f QM is very different from that of NM, there is a sudden discontinuity in density at the star core. This density discontinuity gives rise to a shock at the core, which then propagates outwards with time. This shock is strong enough to combust NM to 2-f QM and mimics a PT in the star. The 2-f QM is not stable, and there is an excess of d quarks. The 2-f matter settles to a stable 3-f matter in the weak timescale.

The study of the conversion of 2-f matter to 3-f matter involves solving a DE governed by the decay of down quarks to strange quarks and the diffusion of quarks across the front boundary. The PT (velocity and time) depends strongly on the temperature of the star, and a cold NS (temperature of $10^{-2}$ MeV) converts quickly than a hotter star (temperature of 1 MeV). Therefore, the GW amplitude and frequency originating from such PT also depends strongly on the star's temperature. For a cold star, the PT takes a few ms to occur, and therefore the frequency is in the 10 kHz range. The GW amplitude of such a star located at 1 Mpc is about $10^{-21}$. For a relatively hotter star (temperature of 0.1 MeV), the signal's frequency is about 1 kHz, and the GW amplitude is about $10^{-23}$. Such signals are more likely to be detected even with the present detectors like aLIGO and VIRGO. For a much hotter star (1 MeV), the frequency is about 100 of Hz and amplitude is about $10^{-25}$. Although the frequency is very much detectable with the present detectors, the amplitude is a challenge for present detectors but likely to be detectable with future detectors like ET.
The GW signal duration is seen to be directly proportional to the temperature of the 2-f star, whereas the GW signal amplitude is inversely proportional to the temperature of the 2-f star. The faster PT results in a stronger GW signal. Hence GW signals from cooler 2-f stars will be detectable even for large source distances, and also, their frequency will be less.

However, if such PT is happening in some galactic pulsars (like Crab pulsar), they are very likely to be detectable with present GW detectors where even hot stars' amplitude is within their detectability range. Detection of such GW signals would validate the existence of quark stars and also the process of PT. We should mention that we have done the calculation for cold NSs, which undergo sudden density fluctuation due to accretion and glitch. The PT scenario is more likely to happen at the end of a supernova in the proto-neutron star or hypermassive NS formed after the binary merging of NSs. Our future aim is to calculate GW signatures from such scenarios.

\section{acknowledgements}
The author RM is grateful to the SERB, Govt. of India for monetary support in the form of Ramanujan Fellowship (SB/S2/RJN-061/2015). RP would like to acknowledge the financial support in the form of INSPIRE fellowship provided by DST, India. SS, RM, and RP would also like to thank IISER Bhopal for providing all the research and infrastructure facilities.

\section*{Data Availability}
The data for this paper is there in the supplementary materials.

	\end{document}